# Electrically programmable chiral MEMS photonics

Longqing Cong[1,2], Prakash Pitchappa[1,2], Nan Wang[3] and Ranjan Singh[1,2,*]

[1]Division of Physics and Applied Physics, School of Physical and Mathematical Sciences, Nanyang Technological University, Singapore 637371, Singapore

[2]Centre for Disruptive Photonic Technologies, The Photonics Institute, Nanyang Technological University, 50 Nanyang Avenue, Singapore 639798, Singapore

[3]Institute of Microelectronics, 11 Science Park Road, 117685, Singapore

[*]E-mail: ranjans@ntu.edu.sg

Abstract: Optical chirality is central to many industrial photonic technologies including enantiomer identification, ellipsometry-based tomography and spin multiplexing in optical communication. However, a substantial chiral response requires a typical three-dimensional (3D) constituent, thereby making the paradigm highly complex. Photonic devices integrated with microelectromechanical systems (MEMS) have shown potential for chiral light control by external stimuli, but the stimuli usually demand a destructive dosage. Here, we report a simple synthetic chiral paradigm that is electrically programmable with self-assembled 3D MEMS cantilevers. This paradigm enables four reconfigurable chiral schemes with dextrorotary, levorotary, racemic and achiral conformations. Optical response of reversible chirality and chiral to achiral switch are electrically encoded following an exclusive OR logical operation with dual-channel bias as low as 10 V. Our device demonstrates a route to electrically actuated synthetic chiral platform and serves as a powerful conformation analysis tool for macromolecules in biology, medicine, chemistry and physics. The prototype delivers a new strategy towards reconfigurable stereoselective photonic applications.

Chirality is ubiquitous in nature from human hands to chemical and biological macromolecules, and climbing plants. Intense research efforts have been devoted to the study of chirality to explore pharmacological effects and even unravel the origin of life.[1] In pharmaceutical industry, it is essential to differentiate the handedness of enantiomers, as it is strongly associated to their potency and toxicity.[2] Currently, circular dichroism (CD) is a routine optical approach to measure chirality.[3] However, the CD response of natural enantiomers is extremely weak, typically in the range of milli-degrees and hence requires large volume (sub-millilitre) of analyte and long integration time (~30 min) to precisely resolve the chirality.[4] Several effective solutions have been reported to overcome the detection limit of naturally weak CD response by incorporating with chiral metamaterials.[5-9] The artificial chiral response as a background amplifies the synthetic CD of the enantiomers to several degrees, leading to a better detection sensitivity of enantiomer handedness. Such an ingenious sensing scheme could be further optimized with improved accuracy by utilizing a real-time reversible background chirality platform that enables a reciprocal handedness detection. The chirality reversal is usually realized by 3D structural reconfiguration of metamaterial constituents via various approaches.[10] Reconfigurable chirality is significant not only for the determination of enantioselective activity in biochemical reactions, but also for exceptional light-manipulation capabilities, including polarization control and detection,[10-12] negative refractive index,[13] perfect lens,[14] and color-tunable polarizer.[15]

Several approaches have been reported for reconfigurable chirality with metamaterials,[16-18] such as by integrating semiconductors (silicon),[19] phase change materials ($Ge_2Sb_2Te_5$),[20] graphene[21], magnesium,[18] DNA scaffold[22] and MEMS actuators.[10, 23] However, the reported active schemes usually demand destructive stimuli, like high optical pump fluence, high temperature, chemical reaction or pneumatic control that makes it inapplicable for bioassay and optical communication systems. Electrostatically actuated bimorph cantilevers in MEMS have exhibited excellent performance in a plethora of interesting applications in reconfigurable photonic devices.[24-32] However, it is still challenging to achieve structural 3D reconfiguration of chiral constituents with weak electrical stimulus.[10] Since chirality is inherently a 3D phenomenon that requires coupling between the local electric and magnetic dipoles,[33, 34] artificial constituents were usually designed with complex conformations,[19] such as spiral with multiple helices,[10, 23, 33] or near-field coupled multilayer metamaterials.[35, 36] So far, small dosage of external stimuli failed to realize the structural reconfiguration of the 3D constituent for chirality reversal.[10, 19]

In this article, we report an electrically programmable chiral paradigm with MEMS cantilevers that enables reversible chirality actuated by *weak stimulus* in the terahertz regime. Two 3D microhelices with opposite handedness form a supercell that are electrically isolated and electrostatically actuated for the dual-channel independent reconfiguration with bias of 10 V. The intrinsic residual stress in the constituent material layers of bimorph cantilever enables the self-assembly of 3D microhelices. This reconfigurable scheme offers *dextrorotary* (*D*), *levorotary* (*L*), *racemic* (*r*) and *achiral* (*a*) conformations, which enables chiral optical response following exclusive OR logical operation.

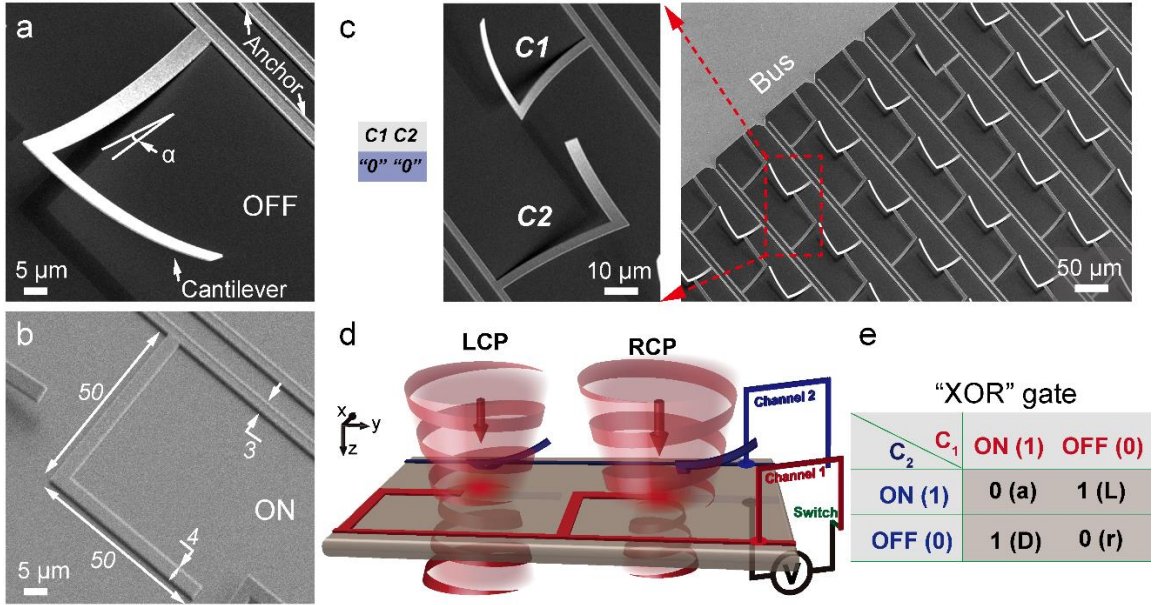

**Figure 1 | Electrically programmable chiral MEMS device.** (a, b) SEM images of one MEMS microhelix at OFF (cantilever suspended) and ON (cantilever actuated) states, respectively. Geometrical parameters of anchor and cantilever are indicated, and the thickness of metal (aluminum, $t_{Al}$) in the bimorph cantilever is 0.4 μm. (c) Scheme of the programmable microhelix array. A zoomed-in image of one microhelix pair with opposite handedness is shown that is electrically isolated and separately actuated via dual channels (C1 and C2). Inputs "0" and "1" correspond to "OFF" and "ON" of external bias, respectively. (d) Schematic diagram of operation as a chirality switch. The programmable chirality switch with [10] input offers dextrorotary (*D-*) conformation that reveals different optical responses for RCP and LCP incidence. (e) Summary of logical operation for the programmable chirality switch in terms of optical chirality output following an "exclusive OR (XOR)" gate. Four different encoding sequences correspond to four chiral states with dextrorotary (*D*), levoratory (*L*), racemic (*r*), and achiral (*a*) conformations.

Here, we utilize an L-shape bimorph cantilever that self-assembles as a 3D microhelix after the release process (see Fig. 1a, also see Methods and supplementary note 1).[25] Such a microhelix possesses intrinsic chirality determined by its handedness, which vanishes when it is electrostatically pulled down onto the substrate forming a planar resonator. A straightforward way to reconfigure the microhelix is by applying a pull-in voltage across the metal (Al) and silicon substrate (10 V, see the supplementary *video*). The electrostatic force neutralizes the residual stress, and pulls the cantilever back onto the substrate surface as shown in Fig. 1b. In order to obtain reversible chirality, metamaterial supercell comprising of two microhelices with opposite handedness was conceived as shown in Fig. 1c. The two neighboring microhelices were electrically isolated from each other, which allowed for the independent control of *D*- and *L*-microhelices through two separate actuation channels (C1 and C2). From the aspect of binary electrical encoding, the state of microhelix without electrical input is defined as "0" (OFF), and 2D planar state is defined as "1" (ON). The independent control of dual-channel electrical inputs [C1 C2] enables four possible reconfiguration states of the metamaterial with unique chiral response at each of those states. With [11] input, all cantilevers are in-plane resulting in a 2D metasurface, which exhibits *achiral* response.[37] For input of [00], mixture of equal-quantity microhelices with opposite handedness exhibits *racemic* conformation. When the handedness based symmetry is broken in the supercell, i.e. for input of [10] (*D*-conformation) or [01] (*L*-conformation), strong intrinsic chirality is observed. The chiral response with [10] input is schematically illustrated in Fig. 1d. The *D*-microhelix in the supercell couples strongly with right-handed circularly polarized (RCP) resulting in a weak transmission intensity. On the other hand, the left-handed circularly polarized light (LCP) couples weakly due to the absence of *L*-microhelix, thereby giving rise to a large differential transmission intensity for RCP and LCP light. This difference generates circular dichroism, which is a direct quantitative evidence of intrinsic chirality. The logical operation of the chirality switch is summarized in Fig. 1e, and it follows exclusive OR (XOR) gate in terms of a dual-channel electrical input and binary chiral optical output. The far-field chiral response is considered as "0" for achiral and racemic conformations with a trivial CD value, and "1" for chiral response with a nontrivial CD value for both *D*- and *L*-conformations.

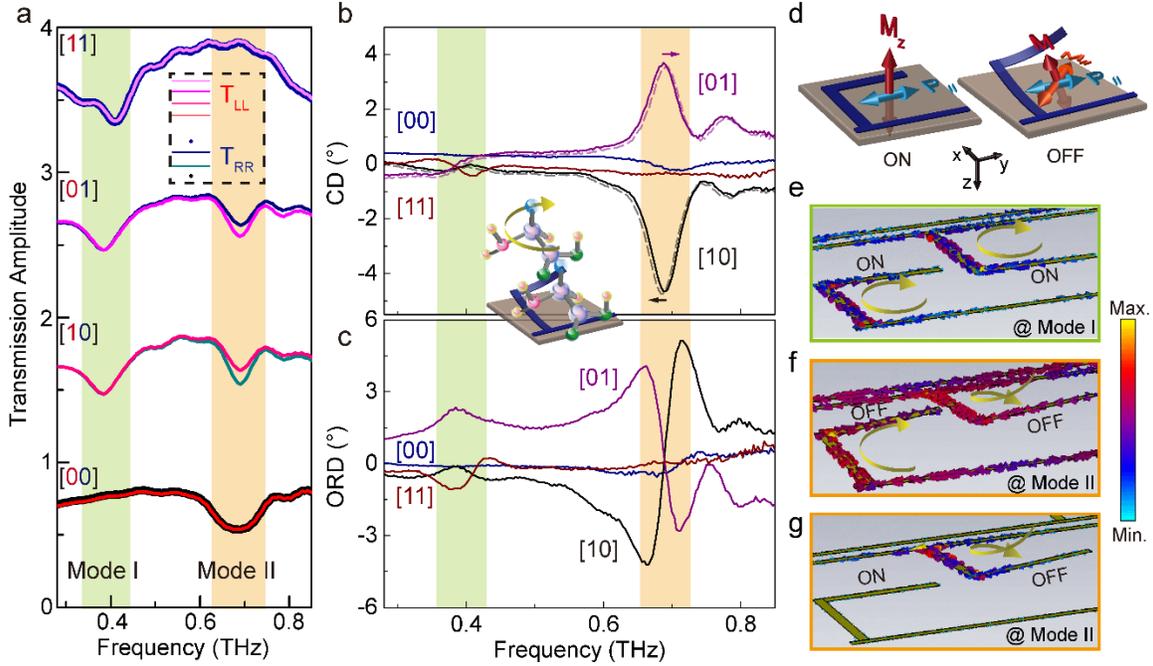

**Figure 2 | Chiral optical response.** (a) Measured co-polarized transmission amplitude of RCP and LCP incidence at four states of the chirality switch. The RCP and LCP spectra reveal identical responses for [00] and [11] inputs, and spectral difference emerges at Mode II with [01] and [10] inputs. (b, c) Circular dichroism and optical rotatory dispersion spectra. The strong CD response would enable a chirality amplification platform for handedness detection of enantiomers. The dashed lines indicate the theoretical spectral response with analyte of *L*-enantiomer of amino acids. The reversible background CD spectra provide a reciprocal detection scheme with opposite spectral frequency shift for identical analyte. (d) Schematic diagram showing the electromagnetic origin of intrinsic chirality. (e) Surface current distribution with [11] input at Mode I. (f, g) Surface current distributions at Mode II with [00] and [10] inputs.

The programmable chiral switching performance in the MEMS metamaterial is experimentally demonstrated using a terahertz time domain spectroscopy system (supplementary Fig. 1 and note 2). In terms of *circular polarization*, the co-polarized transmission spectra of RCP and LCP were measured (Fig. 2a, also see Methods) at four reconfiguration states. With inputs of [00] and [11], the co-polarized transmission spectra ($T_{RR}$ and $T_{LL}$) remain identical. Although achiral response is measured in the far field at these two states, they reveal starkly different underlying mechanism. First, achiral response originates from different spatial conformations: *planar* conformation without intrinsic chirality ([11]) and *3D* conformation of mixed microhelices with opposite handedness for racemic response ([00]). Second, the transmission spectra exhibit large resonance frequency shift from 0.4 THz (mode I) to 0.7 THz (mode II) from [11] to [00] input. The resonance frequency shift is caused by the modulation of effective capacitance between 3D deformation and planar resonators in the metamaterial.[25] For [10] and [01] inputs, both mode I and mode II are simultaneously excited in $T_{RR}$ and $T_{LL}$ spectra, however, spectral difference only emerges in the vicinity of mode II originating from the 3D conformation of microhelix. The transmission difference indicates different complex refractive indices for RCP and LCP incident terahertz waves.[13] The difference in the imaginary part of refractive indices generates circular dichroism (CD), while the difference in the real part of refractive indices manifests as optical rotatory dispersion (ORD).[16] The CD and ORD are defined through transmission spectra as $CD = \tan^{-1}\left[\left(|T_{RR}| - |T_{LL}|\right)\big/\left(|T_{RR}| + |T_{LL}|\right)\right]$ and $ORD = \arg(T_{RR}) - \arg(T_{LL})$, respectively. The measured CD and ORD spectra of the artificial MEMS chirality switch are summarized in Figs. 2b and 2c, respectively. The maximum CD value was measured to be -4.6 ° with *D*-conformation, which is orders of magnitude higher than CD of natural *D*-enantiomers. The CD spectrum is completely reversed by reconfiguring the metamaterial to *L*-conformation with [01] input as shown in Fig 2b. The proposed scheme enables a perfect chirality reversal with opposite CD peaks at exactly the same frequency that would benefit a plethora of applications such as enhanced reciprocal chiral sensing. Apart from the reversible chirality, the chiral response is completely switched off by inputs of [00] (*r*-conformation) and [11] (*a*-conformation) as shown in Fig. 2b. As per the experimental CD spectral output, the MEMS chirality switch shows interesting programmable property that operates on the principle of an XOR gate with binary encoding. Since CD and ORD spectra are related by Kramers-Kronig's equation, ORD spectra also manifest significant enhancement upto 5.1 ° in the vicinity of mode II as shown in Fig. 2c at *D*-conformation, exhibiting large optical activity.

We theoretically explore the possibility of utilizing the proposed scheme for enhanced synthetic CD sensing application. The enantiomers and isomers could be precisely resolved by placing the analyte in the vicinity of the enhanced near-field of the metamaterial microhelices. The synthetic chiral output CD$_s$ with loaded analyte is estimated as[5, 6]:

$$CD_s = CD_b + 4\,pkw\,\mathrm{Im}[\kappa_m] \tag{1}$$

by assuming $kw \ll 1$, where CD$_b$ is the background CD value of the chiral platform, $p$ is a constant ( $p = \left( \left| T_{RL} T_{LR} \right|^2 - \left| T_{LL} T_{RR} \right|^2 \right) \Big/ \left[ \left( \left| T_{RL} \right|^2 + \left| T_{RR} \right|^2 \right)^2 + \left( \left| T_{LL} \right|^2 + \left| T_{LR} \right|^2 \right)^2 \right]$ ), $\kappa_m$ is the intrinsic chirality coefficient of the chiral molecules, and $k$ and $w$ are the wavenumber of light in free space and thickness of the film, respectively. According to Eq. 1, synthetic CD$_s$ is more discernible with giant amplitude that is connected to the background CD$_b$ and the imaginary part of chirality coefficient of enantiomer. By probing the frequency shift ($\Delta\omega$) of synthetic CD$_s$ spectrum before and after placing analyte, the properties of the enantiomers could be resolved via

$$\Delta\omega = -mc\frac{4\operatorname{Im}[\kappa_m]w}{CD_b^{(2)}} \tag{2}$$

where $c$ is the speed of light in vacuum, $m$ is a constant and $CD_b^{(2)}$ is the second derivative of $CD_b$ with respective to $k_0$ (wavevector at resonance wavelength).[5, 6] In this context, when the background $CD_b$ spectrum is reversed, frequency shift would be inversed for the identical analyte. As schematically indicated in Fig. 2b, *L*-enantiomer amino acids that reveal siginificant characteristic absorption spectrum in terahertz regime,[38] would lead to the opposite frequency shift of CD spectrum based on *L*-conformation and *D*-conformation of the MEMS chirality sensing platform. Such an enantioselective technique provides a reciprocal sensing strategy, further guaranteeing the accuracy of enantiomer detection. The proposed programmable MEMS chirality switch would be an excellent candidate as a real-time reversible chiral platform for flexible, high-resolution, fast, and accurate sensing applications.

Regarding the underlying mechanism of intrinsic chirality in terms of electromagnetic interactions, the generalized condition is $\vec{p}\cdot\vec{m} \neq 0$, where $\vec{p}$ and $\vec{m}$ are the moments of net electric and magnetic dipoles, respectively. In a 2D metamaterial, electromagnetically induced magnetic dipole is invariably vertical to the substrate surface ($M_\perp$) with electric dipole in-plane ($P_{//}$) at normal incidence, i.e. $\vec{p}\cdot\vec{m} = 0$ (see Fig. 2d).[37] On the other hand, in the case of 3D microhelices or near-field coupled 3D metamaterials, the net magnetic dipole is no longer vertical to the substrate surface, as the electromagnetic interaction extends spatially in the third dimension as schematically shown in Fig. 2d. For a detailed understanding of the scenario with microhelix, we carried out numerical analysis using finite element method (see Methods) and show the surface current distributions of an individual supercell. At *a*-conformation (Fig. 2e), each L-shape cantilever together with anchor form a 2D split ring resonator, and surface current oscillating on the 2D resonator at Mode I induces a magnetic dipole that is orthogonal to net in-plane electric dipole. As for *r*-conformation in Fig. 2f, similar current loop is observed on an individual microhelix at Mode II, but are guided to flow along the 3D conformation so that $\vec{p} \cdot \vec{m} \neq 0$ in the localized field. However, equal quantity of *D*- and *L*-microhelices results in racemization in the far field. Chiral optical response dominates when the equality of *D*- and *L*-microhelices is broken with [10] and [01] inputs, in which case the electric and magnetic dipoles interact only in *D*- (Fig. 2g) or *L*-microhelices at Mode II.

In the context of electromagnetic understanding, the interaction of electric and magnetic dipoles determines the chiral response. For the MEMS microhelix, the in-plane projection of magnetic dipole ($M_{//}$) depends on the inherent suspension angle of cantilevers ($\alpha$, see Fig. 1a), i.e. the pitch of the microhelix. A larger pitch of microhelix would enable a stronger interaction of electromagnetic components within the scope of current design. For the bimorph MEMS cantilever, the inherent suspension angle originates from the residual stress between Al and Al$_2$O$_3$ layers that is numerically estimated in terms of the radius of curvature ($r$) as follows:

$$\frac{1}{r} = \frac{6n(1+\text{n})(\text{m}\sigma_{Al} - \sigma_d)}{t_{Al}E_{Al}\left[K + 3mn(1+\text{n}^2)\right]} \tag{3}$$

where $K = 1 + 4mn + 6mn^2 + 4mn^3 + m^2n^4$, $m = E_{Al}/E_d$ and $n = t_{Al}/t_d$. Here $t_{Al}$ and $t_d$ are the thickness of metal (Al) and dielectric layer (Al$_2$O$_3$) of the bimorph cantilever; $E_{Al}$ and $E_d$ are the Young's modulus, and $\sigma_{Al}$ and $\sigma_d$ are the residual stress of the Al and $Al_2O_3$, respectively. The Young's modulus and residual stress of materials are constant at room temperature determined by material properties and fabrication process, and thus $r$ can be engineered based on $t_{Al}$ and $t_d$. According to Eq. 3, a smaller $t_{Al}$ would result in a smaller $r$ at constant $t_d$, and thus, a larger pitch of microhelix.

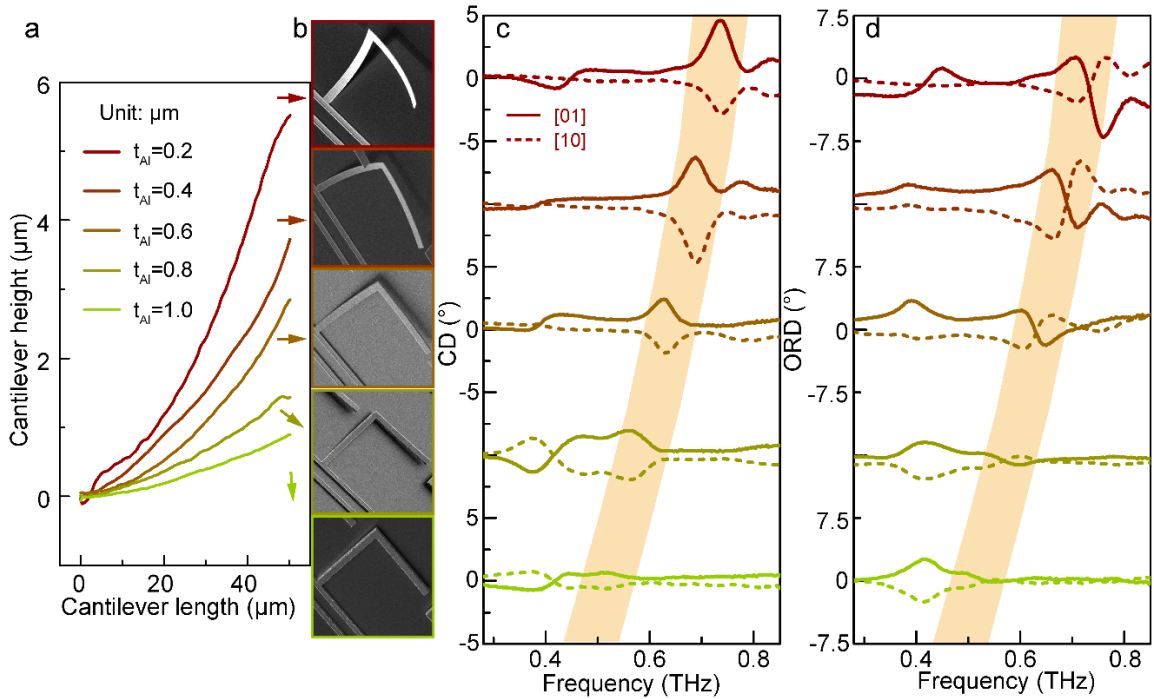

**Figure 3 | Intrinsic chirality modulation at room temperature.** (a) Measured profiles of cantilever arm that is attached to anchor with various metal thicknesses and fixed dielectric layer thickness. (b) SEM images of the microhelix at OFF state. (c, d) CD and ORD spectral response for different samples with *D*- and *L*-conformations. A clear modulation of chirality strength and resonance frequency is observed.

We have experimentally verified the understanding by fabricating real devices with various metal thicknesses of cantilevers ranging from 0.2 to 1.0 μm. The deformation profiles were tested using reflection digital holographic microscope (R-DHM, see supplementary Note

4) and are shown in Fig. 3a. The deformation is clearly larger for thinner metal thickness as shown in the SEM images of each sample at OFF state (Fig. 3b). The chiral optical response is readily tailored by controlling the pitch of microhelix. As revealed in Figs. 3c and 3d, CD and ORD spectra are modulated with a clear trend in the chiral strength as well as resonance frequency at Mode II. The resonance frequency is also altered due to the modulation of effective capacitance in the microhelices at different deformation profiles. Although the pitch of microhelix can also be modulated by changing ambient temperature,[39] extra destructive stimulus of heat or cryogen has to be introduced that increases the system complexity and decreases the feasibility and practicality of the device. Apart from the scalability of optical response for metamaterials, the operating frequency and chirality strength are customizable for on-demand applications with the MEMS technique.

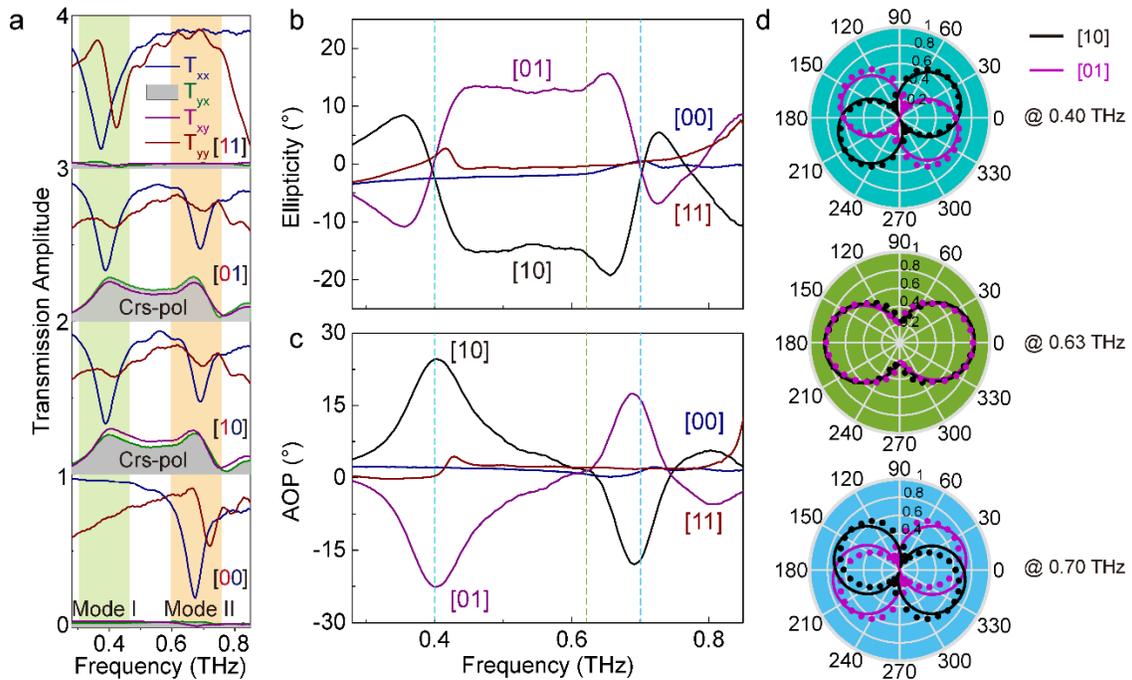

**Figure 4 | Application for dynamic polarization switching.** (a) Measured transmission spectra of four elements in Jones matrix with linearly polarized incidence for four input states of the chirality switch. (b) and (c) The output polarization states described by ellipticity and rotation angle of polarization with *y*-polarized incidence. (d) Visualized output polarization states in polar coordinates at three specific frequencies for [01] and [10] inputs. The output polarization plane is rotated by large angles at 0.4 and 0.7 THz, and polarization state is modulated to elliptical states with opposite handedness at 0.63 THz. Solid lines are polarization states calculated from experimental ellipticity and AOP, and dots are measured angular-resolved transmission amplitude by rotating an analyzer.

One important application of optical chirality is the polarization modulation of light that is a key technology for optical communication, display, and polarimetric spectroscopy. Here, we also explore the programmable polarization modulation capacity of the MEMS chirality switch. To probe the polarization conversion performance, four elements of Jones matrix (see Methods) are measured and plotted with *linearly* polarized incidence at four reconfiguration states of the device with $t_{Al}$ = 0.4 μm, as shown in Fig. 4a. Co-polarized transmission spectra ($T_{xx}$ and $T_{yy}$) reveal different features due to the presence of periodic anchor wires that act as a grating at all four encoding states. Zero cross-polarized components ($T_{yx}$ and $T_{xy}$) were measured with [00] and [11] inputs, indicating no polarization rotation and modulation with *a*- and *r*-conformations. However, large intensity of cross-polarized components emerges with *D*- and *L*-conformations, which gives rise to the modulation of output polarization states. The polarization modulation is numerically expressed in terms of ellipticity and angle of polarization (AOP) as shown in Figs. 4b and 4c. In experiments, the polarization of incident light was aligned parallel to the axis of anchor (defined as *y*-polarized incidence) in order to exclude the anisotropy of anchors. The output polarization state is preserved ("0" output without polarization modulation) for [11] and [00] inputs. For [10] and [01] inputs, strong polarization modulation ("1" output) occurs in a broadband frequency regime extending from 0.4 to 0.7 THz. The output polarization state is modulated from linear to elliptical with opposite handedness, and the ellipticity reaches to ~ ±15 °for [01] and [10] inputs, respectively. In addition, strong optical activity is observed at 0.4 and 0.7 THz, where linear polarization plane is rotated by up to ±25 ° with linear polarization state preserved at 0.4 THz for [10] and [01] inputs, respectively. Cotton effect enables reversed rotation angles at 0.4 and 0.7 THz at a specific conformation.[40]

In order to visualize the output polarization contrast for [10] and [01] inputs (data not shown for [11] and [00] inputs since they are identical to the incident polarization state), polarization states in polar coordinate at three specific frequencies are measured in transmission amplitude at different angles by rotating the analyzer as plotted in Fig. 4d (see supplementary note 2). At 0.4 and 0.7 THz, linear polarization states are preserved, but the polarization planes are rotated by ±25 °and ∓17 °for [10] (*D*-conformation) and [01] (*L*-conformation) inputs, respectively. At 0.63 THz, linear polarization state is transformed to elliptical states with opposite handedness without polarization plane rotation for [10] and [01] inputs (see supplementary Fig. 2). All three polarization conversion patterns enable large output polarization contrast between *D*- and *L*-conformations, which could provide solution to rapid polarization-division multiplexing in high speed optical communication and real-time polarization imaging. In terms of optical output on polarization modulation, the device also enables an XOR gate logical operation demonstrated by the switching behavior between chiral and achiral states. The switching time of the bimorph cantilevers is estimated to be ~2 microseconds as determined by the mechanical resonance frequency (see supplementary Fig. 3 and note 4). Similar to chiral response, the polarization modulation performance including polarization contrast and frequency of operation can be easily tailored by varying the metal thickness of the cantilevers (see supplementary Fig. 4).

In summary, we report a programmable optical chirality switch leveraging on MEMS technique that is electrically actuated with low voltage operation of 10 V. Dual-channel encoding enables four chiral schemes with *D*- and *L*-, *r*-, and *a*-conformations. The proposed paradigm provides potential applications in efficient programmable polarization modulation with significant polarization contrast and rapid encoding speed. This synthetic MEMS chiral platform reveals exceptional flexibility in customizing the operating properties, which would be ideal for precise and rapid screening of enantiomers and isomers, thereby significantly contributing to pharmaceutical industry and development of chirality study in biology, chemistry and physics. Electrically programmable polarization switch would also benefit optical communication for rapid polarization-division multiplexing and imaging. The reconfigurable chiral photonics enabled by MEMS metamaterial with low-bias electrical operation heralds a new paradigm to construct reconfigurable 3D constituents for stereoselective optical and terahertz applications.

**Methods**

*Fabrication*: A CMOS compatible process was adopted for the fabrication of the MEMS chirality switch. An 8″ silicon (Si) wafer was used as the substrate and 100 nm thick silicon oxide ($SiO_2$) was deposited using low pressure chemical vapor deposition (LPCVD) process. This $SiO_2$ layer acts as the sacrificial layer. After the sacrificial layer deposition, the first photolithography step was executed, and subsequently the $SiO_2$ layer at the anchor region was dry etched using reactive ion etching (RIE). Following this, we deposited a 50 nm aluminum oxide ($Al_2O_3$) layer using atomic layer deposition (ALD) process, followed by the sputter deposition of 400 (200, 600, 800, and 1000) nm thick aluminum (Al). Note that at this stage, the bimorph layers ($Al/Al_2O_3$) at the anchor part were in physical contact with Si substrate, and the remaining part was on top of $SiO_2$ layer. Then the next photolithography step was carried out to define the resonator pattern along with the metallic interconnects and bondpads. Subsequently, Al and $Al_2O_3$ layers were both dry etched, leaving the designed L-shaped bimorph pattern. Finally, vapor hydrofluoric acid (VHF) was used to isotropically etch the $SiO_2$ sacrificial layer away, thereby suspending the cantilevers. The initial state of the released cantilevers was bent up due to the residual stress in the bimorph cantilevers, which is defined as "OFF" state. We note that the final VHF release process is not time-controlled which ensures high yield of devices. The mature fabrication procedure produces high quality and large area devices (see Fig. 1 and supplementary *video*).

*Measurements*: A fiber laser based terahertz time-domain spectroscopy system  was used to measure the transmission spectral response of devices and reference in dry nitrogen atmosphere at normal incidence. Fourier transform was carried out to obtain the frequency domain spectra with both amplitude and phase information. The transmission amplitude spectra $|\tilde{T}(\omega)|$ and phase spectra $\varphi(\omega)$ are obtained by $\tilde{T}(\omega) = \tilde{T}^{Sam}(\omega)/\tilde{T}^{Ref}(\omega)$ where $\tilde{T}^{Sam}(\omega)$ is the complex transmission signal of sample and $\tilde{T}^{Ref}(\omega)$ is the complex transmission signal of reference, respectively.

*Simulations*: The numerical simulations (surface current distributions) were carried out using a commercially available software (CST Microwave Studio) by a finite-element frequency-domain solver with unit cell boundary conditions. Floquet ports with 18 modes were defined at the incident and receiving ports. Material parameters were obtained from the software library. Cantilevers were modelled as straight lines with a constant suspending angle which is variable depending on the aluminum thickness.

*Calculations*: The transmission coefficients ($T_{ti}$) under the circular polarization base ($T_{RR}$, $T_{LR}$, $T_{RL}$ and $T_{LL}$) are retrieved from linearly polarized coefficients ($T_{xx}$, $T_{yx}$, $T_{xy}$ and $T_{yy}$, see supplementray Fig. 1 and note 2) where subscripts $i$ and $t$ denote the incident and transmitted light in a certain base. The $T$ matrix of a new base is calculated by

$$\bar{t} = \hat{\Lambda}^{-1}\hat{T}\hat{\Lambda}\bar{i} = \begin{pmatrix} T_{11} & T_{12} \\ T_{21} & T_{22} \end{pmatrix}\begin{pmatrix} \bar{i_1} \\ \bar{i_2} \end{pmatrix}$$ with $\hat{\Lambda}$ being the operator of the basis matrix and $\begin{pmatrix} T_{11} & T_{12} \\ T_{21} & T_{22} \end{pmatrix}$ the

Jones matrix. Under the circular polarization base, the operator is $\hat{\Lambda} = \dfrac{1}{\sqrt{2}}\begin{pmatrix} 1 & 1 \\ i & -i \end{pmatrix}$, and

therefore transmission coefficients are described under the linear polarization base as follows:

$$\hat{T}^f_{circ} = \begin{pmatrix} T_{RR} & T_{RL} \\ T_{LR} & T_{LL} \end{pmatrix} = \frac{1}{2}\begin{pmatrix} (T_{xx}+T_{yy})+i(T_{yx}-T_{xy}) & (T_{xx}-T_{yy})-i(T_{xy}+T_{yx}) \\ (T_{xx}-T_{yy})+i(T_{xy}+T_{yx}) & (T_{xx}+T_{yy})-i(T_{yx}-T_{xy}) \end{pmatrix}$$

where $R$ and $L$ represent RCP and LCP, respectively.

Stokes parameters were introduced to numerically describe the output polarization states with $y$-polarized incidence as following: $S_0 = |T_{xy}|^2 + |T_{yy}|^2$ , $S_1 = |T_{xy}|^2 - |T_{yy}|^2$ , $S_2 = 2|T_{xy}||T_{yy}|\cos\delta$ and $S_3 = 2|T_{xy}||T_{yy}|\sin\delta$ , where $\delta$ is the phase difference ( $\delta = \varphi_{yy} - \varphi_{xy}$ ). The polarization ellipse is calculated by $\tan 2\psi = S_2/S_1$ and $\sin 2\chi = S_3/S_0$ , where $\psi$ and $\chi$ represent angle of polarization ellipse (AOP) and ellipticity, respectively. The ellipticity $\chi = 45°$ indicates a perfect RCP light, and $\chi = -45°$ indicates a perfect LCP light, and $-45° < \chi < 45°$ represents elliptically ($\chi = 0°$, linearly) polarized light.

**Acknowledgements**

The authors acknowledge the research funding support from the Singapore Ministry of Education (MOE2017-T2-1-110, and MOE2016-T3-1-006(S)) and the National Research Foundation (NRF), Singapore and Agence Nationale de la Recherche (ANR), France (grant No. NRF2016-NRF-ANR004).

**Author contributions**

L. C. and R. S. conceived the idea of chirality switch. L. C. designed the experiment, performed the terahertz measurements and simulations. P. P. and N. W. fabricated and characterized the cantilever devices. L. C., P. P. and R. S. wrote the manuscript. R. S. supervised the overall project.

**Competing financial interests**

The authors declare no competing financial interests.

**References:**

1. U. Meierhenrich, *Amino acids and the asymmetry of life: caught in the act of formation*. (Springer Science & Business Media, 2008).

2. A. Hutt and S. Tan, Drugs **52** (5), 1-12 (1996).

3. B. Nordén, *Circular dichroism and linear dichroism*. (Oxford University Press, USA, 1997).

4. S. M. Kelly, T. J. Jess and N. C. Price, Biochimica et Biophysica Acta (BBA)-Proteins and Proteomics **1751** (2), 119-139 (2005).

5. E. Hendry, T. Carpy, J. Johnston, M. Popland, R. Mikhaylovskiy, A. Lapthorn, S. Kelly, L. Barron, N. Gadegaard and M. Kadodwala, Nature nanotechnology **5** (11), 783 (2010).

6. Y. Zhao, A. N. Askarpour, L. Sun, J. Shi, X. Li and A. Alu, Nat Commun **8**, 14180 (2017).

7. A. S. Karimullah, C. Jack, R. Tullius, V. M. Rotello, G. Cooke, N. Gadegaard, L. D. Barron and M. Kadodwala, Adv. Mater. **27** (37), 5610-5616 (2015).

8. M. Matuschek, D. P. Singh, H.-H. Jeong, M. Nesterov, T. Weiss, P. Fischer, F. Neubrech and N. Liu, Small **14** (7), 1702990-n/a (2018).

9. W. Ma, H. Kuang, L. Xu, L. Ding, C. Xu, L. Wang and N. A. Kotov, Nature Communications **4**, 2689 (2013).

10. T. Kan, A. Isozaki, N. Kanda, N. Nemoto, K. Konishi, H. Takahashi, M. Kuwata-Gonokami, K. Matsumoto and I. Shimoyama, Nature Communications **6**, 8422 (2015).

11. Y. Yang, R. C. da Costa, M. J. Fuchter and A. J. Campbell, Nature Photonics **7**, 634 (2013).

12. M. D. Turner, M. Saba, Q. Zhang, B. P. Cumming, G. E. Schröder-Turk and M. Gu, Nature Photonics **7**, 801 (2013).

13. S. Zhang, Y.-S. Park, J. Li, X. Lu, W. Zhang and X. Zhang, Phys. Rev. Lett. **102** (2), 023901 (2009).

14. J. B. Pendry, Phys. Rev. Lett. **85** (18), 3966-3969 (2000).

15. H.-E. Lee, H.-Y. Ahn, J. Mun, Y. Y. Lee, M. Kim, N. H. Cho, K. Chang, W. S. Kim, J. Rho and K. T. Nam, Nature **556** (7701), 360-365 (2018).

16. M. Hentschel, M. Schäferling, X. Duan, H. Giessen and N. Liu, Science advances **3** (5), e1602735 (2017).

17. A. Kuzyk, M. J. Urban, A. Idili, F. Ricci and N. Liu, Science advances **3** (4), e1602803 (2017).

18. X. Duan, S. Kamin, F. Sterl, H. Giessen and N. Liu, Nano Lett. **16** (2), 1462-1466 (2016).

19. S. Zhang, J. Zhou, Y. S. Park, J. Rho, R. Singh, S. Nam, A. K. Azad, H. T. Chen, X. Yin, A. J. Taylor and X. Zhang, Nat Commun **3**, 942 (2012).

20. X. Yin, M. Schäferling, A.-K. U. Michel, A. Tittl, M. Wuttig, T. Taubner and H. Giessen, Nano Lett. **15** (7), 4255-4260 (2015).

21. T.-T. Kim, S. S. Oh, H.-D. Kim, H. S. Park, O. Hess, B. Min and S. Zhang, Science advances **3** (9), e1701377 (2017).

22. R. Schreiber, N. Luong, Z. Fan, A. Kuzyk, P. C. Nickels, T. Zhang, D. M. Smith, B. Yurke, W. Kuang, A. O. Govorov and T. Liedl, Nature Communications **4**, 2948 (2013).

23. T. Kan, A. Isozaki, N. Kanda, N. Nemoto, K. Konishi, M. Kuwata-Gonokami, K. Matsumoto and I. Shimoyama, Appl. Phys. Lett. **102** (22), 221906 (2013).
24. P. Pitchappa, C. P. Ho, L. Cong, R. Singh, N. Singh and C. Lee, Advanced Optical Materials **4** (3), 391-398 (2016).
25. P. Pitchappa, H. Chong Pei, L. Dhakar, Q. You, N. Singh and L. Chengkuo, Microelectromechanical Systems, Journal of **24** (3), 525-527 (2015).
26. Z. Han, K. Kohno, H. Fujita, K. Hirakawa and H. Toshiyoshi, Opt. Express **22** (18), 21326-21339 (2014).
27. A. Isozaki, T. Kan, H. Takahashi, K. Matsumoto and I. Shimoyama, Opt. Express **23** (20), 26243-26251 (2015).
28. X. Liu and W. J. Padilla, Advanced Optical Materials **1** (8), 559-562 (2013).
29. W. M. Zhu, A. Q. Liu, T. Bourouina, D. P. Tsai, J. H. Teng, X. H. Zhang, G. Q. Lo, D. L. Kwong and N. I. Zheludev, Nat Commun **3**, 1274 (2012).
30. X. Liu and W. J. Padilla, Optica **4** (4), 430-433 (2017).
31. T. Roy, S. Zhang, I. W. Jung, M. Troccoli, F. Capasso and D. Lopez, APL Photonics **3** (2), 021302 (2018).
32. E. Arbabi, A. Arbabi, S. M. Kamali, Y. Horie, M. Faraji-Dana and A. Faraon, Nature Communications **9** (1), 812 (2018).
33. J. K. Gansel, M. Thiel, M. S. Rill, M. Decker, K. Bade, V. Saile, G. von Freymann, S. Linden and M. Wegener, Science **325** (5947), 1513-1515 (2009).
34. A. Y. Zhu, W. T. Chen, A. Zaidi, Y.-W. Huang, M. Khorasaninejad, V. Sanjeev, C.-W. Qiu and F. Capasso, Light: Science & Applications **7** (2), 17158 (2018).
35. Y. Zhao, M. Belkin and A. Alù, Nat Commun **3**, 870 (2012).
36. Y. Ye and S. He, Appl. Phys. Lett. **96** (20), 203501 (2010).
37. E. Plum, X. X. Liu, V. A. Fedotov, Y. Chen, D. P. Tsai and N. I. Zheludev, Phys. Rev. Lett. **102** (11), 113902 (2009).
38. M. Yamaguchi, F. Miyamaru, K. Yamamoto, M. Tani and M. Hangyo, Appl. Phys. Lett. **86** (5), 053903 (2005).
39. P. Pitchappa, M. Manjappa, H. N. S. Krishnamoorthy, Y. Chang, C. Lee and R. Singh, Appl. Phys. Lett. **111** (26), 261101 (2017).
40. E. L. Eliel and S. H. Wilen, *Stereochemistry of organic compounds*. (John Wiley & Sons, 2008).